\documentstyle[12pt]{article}

\begin{document}

\begin{flushright}
DESY 95--245 \\
SLAC--PUB--95--7085\\
RU--95--93 \\
AZPH--TH--95/27
\end{flushright}

\begin{center}
\bigskip\bigskip
{\Large \bf
The Generalized Crewther Relation in QCD and its Experimental
Consequences}
      \footnote{\baselineskip=12pt
Work partially supported by the Department of Energy,
contracts DE--AC03--76SF00515 and DE--FG02--93ER--40792; Project N
96-02-18897-a submitted to the Russian Fund for Fundamental
Research; and ISF grant  N6J000 and ICFPM graduate student
fellowship, grant 932492.}

\vspace{0.2in}

{\bf S. J. BRODSKY}

\vspace{0.1in}

{\baselineskip=14pt
Stanford Linear Accelerator Center \\
Stanford University, Stanford, California 94309, USA \\}
\vspace{0.2in}

{\bf G. T. GABADADZE}\footnote{\baselineskip=12pt
On leave of absence from Institute
for Nuclear Research of the Russian Academy of Sciences, Moscow
117312, Russia.}
\vspace{0.2in}

{\baselineskip=14pt
Department of Physics and Astronomy, Rutgers University \\
Piscataway, New Jersey 08855, USA}
\vspace{0.2in}

{\bf  A. L. KATAEV}
\vspace{0.1in}

{\baselineskip=14pt
Institute for Nuclear Research of the Russian Academy
of Sciences, \\
Moscow 117312, Russia\\}
\vspace{0.2in}

{\bf H. J. LU}
\vspace{0.1in}

{\baselineskip=14pt
Department of Physics, University of Arizona\\
Tucson, Arizona 85721, USA}
\end{center}
\vspace{.5cm}
\begin{center}
Submitted to Physics Letters B.
\end{center}
\vspace{.75cm}
\newpage

\vspace{0.9cm}
\begin{center}
{\bf Abstract}
\end{center}

We use the BLM scale-fixing prescription to derive a
renormalization-scheme invariant relation between the coefficient
function for the Bjorken sum rule for polarized deep inelastic
scattering and the $R$-ratio for the $e^+e^-$ annihilation cross
section. This relation provides a generalization of the Crewther
relation to non-conformally invariant gauge theories. The derived
relations allow one to calculate unambiguously without
renormalization scale or scheme ambiguity the effective charges of
the polarized Bjorken and the Gross-Llewellen Smith sum rules from
the experimental value for the effective charge associated with
$R$-ratio. Present data are consistent with the generalized Crewther
relations, but measurements at higher precision and energies will be
needed to decisively test these fundamental relations in QCD.

\bigskip\bigskip

\leftline{{\bf 1. Introduction}}
\bigskip

In 1972 Crewther \cite{Crewther} derived a  remarkable consequence
of the operator product expansion for conformally invariant gauge
theory.  Crewther's relation has the form \begin{equation} 3 S = K
R', \end{equation} where $S$ is the value of the anomaly controlling
$\pi^0 \to \gamma \gamma$ decay, $K$ is the value of the Bjorken sum
rule in polarized deep inelastic scattering, and $R'$ is the
isovector part of the annihilation cross section ratio $\sigma(e^+
e^- \to $hadrons)/$\sigma(e^+ e^- \to \mu^+ \mu^-)$.

The status of  Crewther's relation within perturbative QCD where
conformal invariance does not hold was recently analyzed in Ref.
\cite{BK}. Using the existing multi-loop calculations for the
coefficient function for $R(s)=\sigma_{\rm tot}(e^+e^-
\rightarrow{\rm hadrons})$ \cite{ChKT,GKL,SS} and the polarized
Bjorken sum rule \cite{GorLar,LarVerm}, the  authors of Ref.
\cite{BK} observed that all perturbative corrections of the type
$C_{\rm F}^N \alpha_{\rm s}^N,~~1\leq N \leq 3 $ are mutually
cancelled in the product of the coefficient function for the
polarized Bjorken sum rule and the Adler's function for $e^+e^-$
annihilation into hadrons. They also showed that the surviving
corrections are suitably grouped in the two-loop QCD
$\beta$-function and derived the QCD generalization of the Crewther
relation at the three-loop order.  It has been demonstrated in Ref.
\cite{GK} that the cancellation of the $C_{\rm F}^N \alpha_{\rm s}^N
$ type corrections (for arbitrary $N$) is related to the
nonrenormalizability of the axial-vector-vector three-point Green
function occurring when the appropriate normalization for the
non-singlet axial current is chosen \cite{AdlerBardeen,Schrier} (for
the discussions of this subject see e.g. the works from Ref.
\cite{GabPiv}). On the other hand, the authors of Ref. \cite{BLB}
have proposed to resolve the problem of the renormalization scale
ambiguity by focusing on relations between experimentally measurable
observables and using the BLM prescription \cite{BLM}. In the
present paper this idea is applied to the Crewther relation in QCD.

A helpful tool for relating physical quantities is the ``effective
charge" approach.  Any perturbatively calculable physical quantity
can be used to define an effective charge
\cite{Grunberg,DharGupta,GuptaShirkovTarasov} by incorporating the
entire radiative correction into its definition.  An important
result is that all effective charges $\alpha_A(Q^2)$ satisfy the
renormalization group equation with the same first ($\beta_0$) and
the second ($\beta_1$) coefficients of the $\beta$ function. The
renormalization group evolutions of the effective charges only
differ through the third and higher coefficients of the effective
$\beta$ functions, which  are process-dependent but
scheme-independent. Thus, any effective charge can be used as a
reference coupling of any renormalization scheme, including the
$\overline{\rm MS}$-scheme. Each effective charge $\alpha_A(Q^2)$ is
a special case of the universal coupling function $\alpha(Q^2,
\beta_n)$, $n\geq 2$ (see e.g.  \cite{BrodskyLuStuecPeter}; for the
recent theoretical studies of the ``scheme-invariant'' expansions
see Ref. \cite{DSlavnov}).

For example, consider the Adler function \cite {Adler} for the $e^+
e^-$ annihilation cross section
\begin{equation}
D(Q^2)=-12\pi^2 Q^2{d\over dQ^2}\Pi(Q^2),~
\Pi(Q^2) =-{Q^2\over 12\pi^2}\int_{4m_{\pi}^2}^{\infty}{R_{e^+
e^-}(s)ds\over s(s+Q^2)}.
\end{equation}
The entire radiative correction to this function is defined
as the effective charge
$\alpha_D(Q^2)$ :
\begin{eqnarray}
    D \left( Q^2/ \mu^2, \alpha_{\rm s}(\mu^2/
\Lambda^2_{\overline{\rm MS}}) \right)
&=& D \left (1, \alpha_{\rm s}(Q^2/
\Lambda^2_{\overline {\rm MS}})\right) \equiv \\
\label{3}
    3 \sum_f Q_f^2 \left[ 1+ {\alpha_D(Q^2/\Lambda_D^2) \over \pi}
                   \right]
    +( \sum_f Q_f)^2C_{\rm L}(Q^2)
&\equiv& 3 \sum_f Q_f^2 C_D(Q^2)+( \sum_f Q_f)^2C_{\rm L}(Q^2),
\nonumber
\end{eqnarray}
where $\Lambda_D$ is the scheme-independent effective scale
parameter.  The coefficient $C_{\rm L}(Q^2)$ appears at the third
order in perturbation theory and is related to the ``light-by-light
scattering type" diagrams.  (Hereafter $\alpha_{\rm s}$ will denote
the ${\overline{\rm MS}}$ scheme strong coupling constant).

We can similarly define the entire radiative correction to the
Bjorken sum rule as the effective charge $\alpha_{g_1}(Q^2)$ where
$Q$ is the corresponding momentum transfer:\footnote{\baselineskip
12pt For a detailed analysis of the available experimental data for
this sum rule, see Ref.  \cite{EllisKarliner}.}
\begin{equation}
\int_0^1 d x \left[
g_1^{ep}(x,Q^2) - g_1^{en}(x,Q^2) \right]
   \equiv {1\over 6} \left|g_A \over g_V \right|
   C_{\rm Bj}(Q^2)
 = {1\over 6} \left|g_A
\over g_V \right| \left[ 1- {\alpha_{g_1}(Q^2/\Lambda_{g_1}^2) \over
   \pi} \right]  .
\end{equation}

It is straightforward to algebraically relate $\alpha_{g_1}(Q^2)$ to
$\alpha_D(Q^2)$ using the known expressions to three loops in the
$\overline{\rm MS}$ scheme. If one chooses new renormalization
scales to re-sum all quark and gluon vacuum polarization corrections
into $\alpha_D(Q^2)$,  then the final result turns out to be
remarkably simple and can be expressed in the following form
\cite{BLB}:
\begin{equation}
\widehat{\alpha}_{g_1} (Q)=\widehat{\alpha}_D (Q^*)-\widehat{\alpha}_D^2
 (Q^{**})
+\widehat{\alpha}_D^3 (Q^{***})+ \cdots,
\label{EqCrewtherSeries}
\end{equation}
where $ \widehat{\alpha}_{g_1} (Q)=\frac{3 C_{\rm F}}{4 \pi}
\alpha_{g_1} (Q)$, $\widehat{\alpha}_D (Q)=\frac{3 C_{\rm F}}{4 \pi}
\alpha_D (Q)$. This ``commensurate scale relation" (CSR) was derived
in Ref. \cite{BLB} by using distinct commensurate scales $Q^*$,
$Q^{**}$ and $Q^{***}$.

These commensurate scales were related in Ref. \cite{BLB} through
the mean value theorem to the mean virtuality of the momenta of the
gluon propagators which appear in each respective order of the
perturbation theory. (For discussions and various applications of
this physical language, see Refs. \cite{BLB,BLM,LM,Neubert,BBB}.)

In this paper we will show that it is also possible and convenient
to choose one unique mean scale $\overline Q^*$ in $\alpha_D(Q)$ so
that the corrections at the right-hand side will also reproduce the
coefficients of the geometric progression. The possibility of using
a single scale in the generalization of the BLM prescription beyond
the next-to-leading order (NLO) was first considered in Ref.
\cite{GrunKat}. The new single-scale Crewther relation has the form:
\begin{equation}
\widehat{\alpha}_{g_1}(Q)=\widehat{\alpha}_D(\overline Q^*)-
\widehat{\alpha}_D^2(\overline Q^*)+\widehat{\alpha}_D^3(\overline Q^*)
+ \cdots,
\end{equation}

The coefficients in the CSR relating $\alpha_{g_1}(Q)$ to
$\alpha_D(Q)$ (aside for a factor of $C_{\rm F},$ (in QCD $C_{\rm F}=4/3$)
which can be absorbed in the definition of the couplings) are
actually independent of color and are the same in Abelian,
non-Abelian, and conformal gauge theory.  The non-Abelian structure
of the theory is reflected in the expression for the scale
$\overline{Q}^{*}$. Note that the $\overline{\rm MS}$
renormalization scheme is used here for calculational convenience;
it serves simply as an intermediary between observables. This is
equivalent to the group property defined by Peterman and
St\"uckelberg \cite{StueckelbergPeterman} which ensures that the
forms of the CSR relations in perturbative QCD are independent of
the choice of an intermediate renormalization scheme or
renormalization scale $\mu$.  (The renormalization group method was
developed by Gell-Mann and Low \cite{GellMannLow} and by Bogoliubov
and Shirkov \cite{BogoliubovShirkov}.)

\bigskip
\leftline{{\bf 2.
Derivation of Commensurate Scale Relations and the Generalized }}
\leftline{{\bf Crewther Relation}}
\bigskip

Let us now discuss in more detail the derivation of the relation
between observables in QCD. Any effective charge in perturbative QCD
can be written in the following form:

\begin{eqnarray}
\frac{\alpha_1 (Q^2/\Lambda_{1 \rm eff}^2)}{4\pi}
&=&
\frac{\alpha_{\rm s} (Q^2/\Lambda_{\overline{\rm MS}}^2)}{4\pi}
+
\Bigl( A_1 + B_1 \beta_0
\Bigr)
\left( \frac{\alpha_{\rm s} (Q^2/
\Lambda_{\overline{\rm MS}}^2)}{4\pi}\right)^2\cr
&+&
\Bigl( C_1 + D_1 \beta_0 + E_1 \beta_0^2 + B_1 \beta_1
\Bigr)
\left( \frac{\alpha_{\rm s} (Q^2/
\Lambda_{\overline{\rm MS}}^2)}{4\pi}\right)^3+ \cdots ,
\end{eqnarray}
where $\Lambda_{1 \rm eff}^2=\Lambda_{{\overline{\rm MS}}}^2
\exp({A_1+B_1\beta_0\over \beta_0})$ and $\beta_0$, $\beta_1$ are
the scheme-invariant first two coefficients of QCD $\beta$-function
defined as $Q^2d(\alpha_{\rm s}/4\pi)/dQ^2 =\beta(\alpha_{\rm
s})=-\sum_{i\geq0} \beta_i(\alpha_{\rm s} /4\pi)^{i+2}$. Within our
normalization conditions the first two scheme-independent
coefficients of the QCD $\beta$-function read: $\beta_0=11-(2/3)f$,
$\beta_1=102-(38/3)f$. In the MS-like schemes the third coefficient
was calculated in Ref. \cite{Lesha} and has the following form
$\beta_2=2857/2-(5033/18)f+(325/54)f^2$.

Notice that the appropriate adjustment of the $D_1$ multiplier
allows one to obtain the $ B_1$ coefficient in front of $\beta_1$ in
the next-next-to-leading  order (NNLO) (it coincides with the
coefficient in front of $\beta_0$ in the NLO).

Similarly, given a second effective charge $\alpha_2(Q)$, we can put
it in the form
\begin{eqnarray}
\frac{\alpha_2 (Q^2/\Lambda_{2 \rm eff}^2)}{4\pi} &=&
\frac{\alpha_{\rm s} (Q^2/\Lambda_{\overline{\rm MS}}^2)}{4\pi}
+\Bigl( A_2 + B_2 \beta_0\Bigr)
\left( \frac{\alpha_{\rm s} (Q^2/\Lambda_{\overline{\rm MS}}^2)}{4\pi}
\right)^2\cr &+& \Bigl( C_2 + D_2 \beta_0 + E_2 \beta_0^2
       + B_2 \beta_1 \Bigr)
\left( \frac{\alpha_{\rm s} (Q^2/\Lambda_{\overline{\rm MS}}^2)}{4\pi}
\right)^3
+ \cdots ,
\label{EqGeneralFormSecondCharge}
\end{eqnarray}
with $\Lambda_{2 \rm eff}^2=\Lambda_{{\overline{\rm MS}}}^2
\exp({A_2+B_2\beta_0\over \beta_0})$. The two effective charges
$\alpha_1(Q)$ and $\alpha_2(Q)$ are related through the following
series,
\begin{eqnarray}
\frac{\alpha_1 (Q^2/\Lambda_{1 \rm eff}^2)}{4\pi} &=&
 \frac{\alpha_2 (Q^2/\Lambda_{2 \rm eff}^2)}{4\pi}+
\Bigl( A_{12} + B_{12} \beta_0 \Bigr)
\left( \frac{\alpha_2 (Q^2/\Lambda_{2 \rm eff}^2)}{4\pi}
\right)^2 +\cr
&+& \Bigl( C_{12} + D_{12} \beta_0 + E_{12} \beta_0^2
       + B_{12} \beta_1\Bigr)
\left( \frac{\alpha_2 (Q^2/\Lambda_{2 \rm eff}^2)}{4\pi}\right)^3
+ \cdots,
\label{EqGeneralRelationTwoCharges}
\end{eqnarray}
where the coefficients $A_{12}, B_{12}, C_{12}, D_{12}$ and $E_{12}$
are given by: $A_{12} = A_1-A_2$, $B_{12} = B_1-B_2$, $C_{12} =
C_1-C_2-2(A_1-A_2)A_2$, $D_{12} = D_1-D_2-2(A_1 B_2 + A_2 B_1) + 4
A_2 B_2$, $E_{12} = E_1-E_2-2(B_1-B_2)B_2 .$

In the multiple-scale CSR approach derived in Ref. \cite{BLB}, one
absorbs order-by-order the coefficients which depend on the number
of flavors into the redefinition of the commensurate scales. As
discussed in Ref. \cite{BLB}, this feature is analogous to  analyses
in QED, where the skeleton diagrams of different orders can have
different renormalization scales \cite{BLM,HJL}. After absorbing the
running coupling effects order-by-order, the authors of Ref.
\cite{BLB} obtained
\begin{equation}
\frac{\alpha_1 (Q^2/\Lambda_{1 \rm eff}^2)}{4\pi}=
\frac{\alpha_2 (Q^{*2}/\Lambda_{2 \rm eff}^2)}{4\pi}+ A_{12}
\left( \frac{\alpha_2 (Q^{**2}/\Lambda_{2 \rm eff}^2)}{4\pi}
\right)^2+C_{12}
\left( \frac{\alpha_2 (Q^{***2}/\Lambda_{2 \rm eff}^2)}{4\pi}
\right)^3 + \cdots,
\end{equation}
where in the definitions of this work the relations between the
renormalization scales read
\begin{equation}
\ln \left( Q^{*2} /Q^2\right)=-  B_{12} + \beta_0
\left( B_{12}^2-E_{12}\right)
\left( \frac{\alpha_2 (Q^{*2}/\Lambda_{2 \rm eff}^2)}{4\pi}
\right), \ln \left( Q^{**2} /Q^2 \right)=- \frac{D_{12}}{2A_{12}} ,
\end{equation}
and $Q^{***}$ can be chosen as $Q^{**}$ if the NLO coefficient is
non-zero, or as $Q^{*}$ if the NLO coefficient is zero. As discussed
in Ref. \cite{BLB}, the use of different renormalization scales at
different orders of perturbation theory can be related to the
different mean momenta in the skeleton graphs contributing at each
order. Notice also that since the coefficient before $\beta_1$-term
in the NNLO contribution to Eq. (9) is equal to the coefficient
$B_{12}$ of the NLO correction to Eq. (9), the absorption of the
proportional to $\beta_0$ NLO term in Eq. (9) into the scale $Q^*$
automatically leads to the nullification of the NNLO contributions
into Eq. (9), which is  proportional to the $\beta_1$-coefficient of
the QCD $\beta$-function. The remaining NNLO terms in Eq. (9), which
turn out to be proportional to $\beta_0^2$ and $\beta_0$, were
absorbed in Ref. \cite{BLB} into the scales $Q^*$ and $Q^{**}$. An
alternative approach is to use a single renormalization scale for
all orders in the right-hand side of Eq.
(\ref{EqGeneralRelationTwoCharges}). This approach for generalizing
the BLM procedure has been considered in Ref. \cite{GrunKat}. In the
single-scale approach we have
\begin{equation}
\frac{\alpha_1 (Q^2/\Lambda_{1 \rm eff}^2)}{4\pi}=
\frac{\alpha_2 (\overline{Q}^{*2}/\Lambda_{2 \rm eff}^2)}{4\pi}+
 A_{12}
\left( \frac{\alpha_2 (\overline{Q}^{*2}/\Lambda_{2 \rm eff}^2)}{4\pi}
\right)^2+C_{12}
\left( \frac{\alpha_2 (\overline{Q}^{*2}/\Lambda_{2 \rm eff}^2)}{4\pi}
\right)^3 + \cdots,
\end{equation}
where the expansion coefficients in this series are identical to
those of the multiple-scale case in Eq.
(\ref{EqGeneralRelationTwoCharges}). However, there is only one
single renormalization scale
\begin{equation}
\ln \left(\overline{ Q}^{*2} /Q^2 \right)=- B_{12} +
\left[ \beta_0 \  \left( B_{12}^2-E_{12} \right) +
2 A_{12}B_{12} -  D_{12}\right]
\left( \frac{\alpha_2 (\overline{Q}^{*2}/\Lambda_{2 \rm eff}^2)}{4\pi}
\right).
\end{equation}
Hereafter we will simply drop  the dependence of the coupling
constants on the QCD parameters $\Lambda_{\rm eff}$.

We now apply the above general procedures to derive a single-scale
CSR between  the effective couplings for the coefficient function of
Bjorken polarized sum rule and the Adler's function for the $e^+e^-$
annihilation. The perturbative series for $\alpha_{g_1}(Q)/4\pi$
using dimensional regularization and the $\overline{\rm MS}$ scheme
with the renormalization scale fixed at $\mu=Q$ has been computed at
the NNLO in \cite{LarVerm}. The effective charge for the
annihilation cross section has been computed in the $\rm
\overline{\rm MS}$ scheme at the NNLO with the renormalization scale
fixed at $\mu=Q$  in Ref. \cite{GKL,SS}. After eliminating the
$\overline{\rm MS}$-scheme and applying the BLM procedure, the
single scale result has the following simple form:

\begin{equation}
\widehat{\alpha}_{g_1}(Q)=\widehat{\alpha}_D(\overline Q^*)-
\widehat{\alpha}_D^2(\overline Q^*)+\widehat{\alpha}_D^3(\overline Q^*)
+ \cdots,
\label{EqSingleScaleCrewtherSeries}
\end{equation}
where $\widehat{\alpha}_{g_1}=(3 C_{\rm F}/4\pi) \alpha_{g_1}$ and
$\widehat{\alpha}_D=(3 C_{\rm F}/4\pi) \alpha_D$. Recalling that $
\widehat{\alpha}_{g_1}(Q) = C_{\rm Bj}(Q)-1$ , $\widehat{\alpha}_D(\overline
Q^*)= C_{D}(\overline Q^*)-1$ , we get the simple relation
\begin{equation}
C_{\rm Bj}(Q)C_D(\overline Q^*)=1.
\label{EqGeneralizedCrewther}
\end{equation}
Here
\begin{eqnarray}
\ln \left({ {\overline Q}^{*2} \over Q^2} \right) &=&
{7\over 2}-4\zeta(3)+\left(\frac{\alpha_D (\overline Q^*)}{4\pi}
\right)\Biggl[ \left(
            {11\over 12}+{56\over 3} \zeta(3)-16{\zeta^2(3)}
      \right) \beta_0\cr &&
      +{26\over 9}C_{\rm A}
      -{8\over 3}C_{\rm A}\zeta(3)
      -{145\over 18} C_{\rm F}
      -{184\over 3}C_{\rm F}\zeta(3)
      +80C_{\rm F}\zeta(5)
\Biggr].
\label{EqLogScaleRatio}
\end{eqnarray}
where in QCD $C_{\rm A}=3$, $C_{\rm F}=4/3$. The relations
(\ref{EqGeneralizedCrewther}) and (\ref{EqLogScaleRatio}) show how
the coefficient functions for these two different processes are
related to each other at their respective commensurate scales. The
evaluation of one of them at the appropriate physical scale gives us
information about the second one at the different physical scale.
Notice also that all the $\zeta(3)$ and $\zeta(5)$ dependencies have
been absorbed into the renormalization scale $\overline Q^*$. The
explicit forms for the corresponding multiple scales are given in
Ref. \cite{BLB}.

\bigskip
\leftline{{\bf 3. The Generalized Crewther Relation and its
Experimental Consequences }}
\bigskip

The simple form of Eqs. (5), (6), (15) points to the existence of a
``secret symmetry" between $\alpha_D(Q)$ and $\alpha_{g_1}(Q)$ which
is revealed after the application of the BLM scale setting
procedure. In the conformally invariant limit, {\it i.e.}, for
vanishing beta functions, the generalized Crewther relation derived
in Ref. \cite{BK} becomes
\begin{equation}
(1+\widehat{\alpha}_D^{\rm eff})(1-\widehat{\alpha}_{g_1}^{\rm eff})=1,
\end{equation}
which is equivalent to  Eqs. (\ref{EqCrewtherSeries}), (6),
(\ref{EqGeneralizedCrewther}). Thus Eqs. (\ref{EqCrewtherSeries}),
(6)  or (\ref{EqGeneralizedCrewther}) can be regarded as the
extension of the Crewther relation to non-conformally invariant
gauge theory in which all effects of the non-vanishing QCD
$\beta$-function are absorbed into the scale of the coupling and
each physical quantity has its appropriate scale of momenta or
energy.

We can also write down analogous equations for the polarized Bjorken
sum rule or for the Gross-Llewellyn Smith sum rule, defined as
\begin{equation}
\frac{1}{2}\int_0^1 F_3^{\nu p+\overline{\nu}p}(x,Q^2)dx=3
C_{\rm GLS}(Q^2)=3[1-\frac{\alpha_{\rm GLS}(Q^2)}{\pi}]
\end{equation}
and the Adler's function for the $e^+e^-$ annihilation process.
Taking into account the additional three-loop ``light-by-light-type"
diagrams, the needed relations can be expressed in the form

\begin{equation}
{1\over 3\sum_f Q_f^2} D(\overline Q^*)
C_{\rm Bj}(Q)=1+\varepsilon_1(Q)+\cdots,
\end{equation}
and
\begin{equation}
{1\over 3\sum_f Q_f^2} D(\overline Q^*)
C_{\rm GLS}(Q)=1+\varepsilon_2(Q)+\cdots,
\end{equation}
where the dots denote higher order corrections and
$\varepsilon_1(Q),~~\varepsilon_2(Q)$ are related to the
light-by-light scattering type diagrams:

\begin{eqnarray}
         \varepsilon_1(Q)
&\equiv& \left(\frac{44}{9}-\frac{32}{3}\zeta(3)\right)
         \frac{d^{abc}d^{abc}}{C_{\rm F} N_c}
         \frac{\left( \sum_f Q_f\right)^2}{\sum_f Q_f^2}
         \left(\frac{\alpha_D(Q)}{4\pi}
         \right)^3 ,
\cr
         \varepsilon_2(Q)
&\equiv& \left(\frac{44}{9}-\frac{32}{3}\zeta(3)\right)
         \frac{d^{abc}d^{abc}}{C_{\rm F} N_c} \frac{\left( \sum_f
         Q_f\right)^2-f\sum_f Q_f^2}{\sum_f Q_f^2}
         \left(\frac{\alpha_D(Q)}{4\pi}
         \right)^3 .
\end{eqnarray}
The scales $Q$ and $\overline Q^*$ are defined above, $N_c=3$ is the
number of colors, $f$ is the number of active  flavors of quarks
with the charges $Q_f$ and in QCD $d^{abc}d^{abc}=40/3$. Strictly
speaking, the right hand sides  of these equations depend on a scale
variable $\overline Q^{*}$; however recalling that we are not able
to control the fourth order corrections to these relations we just
replace $\overline Q^{*}$ by $Q$ in the expressions for the third
order corrections. Notice that the numerical values of
$\varepsilon_1(Q)$ and $\varepsilon_2(Q)$ terms are very small.

We now address the question whether one can extract
phenomenologically useful consequences from these relations. To do
this we shall express the perturbatively calculated Adler's function
through the experimentally measurable $R$-ratio for the $e^+e^-$
annihilation cross section. As is well known, the perturbatively
defined Adler's function is computed for the space-like transfer
momenta. In order to obtain the expression for the measurable
$R$-ratio one has to analytically continue from the space-like to
the time-like momenta. If one is sufficiently far from the lowest
resonance production thresholds, it is possible to perform this
continuation using the following simple formula

\begin{equation}
R_{e^+e^-}(s)={1\over 2 \pi i}
\int_{-s-i\varepsilon}^{-s+i\varepsilon} {d\tau \over
\tau}D(\alpha_{\rm s}(\tau))
= 3\sum_f
Q_f^2[1+\frac{\alpha_R(\sqrt{s})}{\pi}].
\end{equation}
In general, this procedure results in the appearance of $\pi^2$ like
terms in the coefficient functions.  More precisely, in our case the
following shifts of numerical coefficients of Eq.
(\ref{EqGeneralFormSecondCharge}) and Eq.
(\ref{EqGeneralRelationTwoCharges}) take place: $E_2\rightarrow
E_2-{1\over 3}\pi^2;$ $E_{12}\rightarrow E_{12}+{1\over 3}\pi^2$. In
addition it is necessary also to perform the replacements:
$\overline Q^*\rightarrow \sqrt s$, $\alpha_D\rightarrow\alpha_R$,
where $s$ is defined through the value of $Q^2$
\begin{eqnarray}
\ln \left({ Q^2 \over s} \right)&=&-{7\over 2}+4\zeta(3)-
\left(\frac{\alpha_R (\sqrt{s})}{4\pi}\right)
\Biggl[ \left(
            {11\over 12}+{56\over 3} \zeta(3)-16{\zeta^2(3)}
            -\frac{\pi^2}{3}
      \right) \beta_0\cr &&
      +{26\over 9}C_{\rm A}
      -{8\over 3}C_{\rm A}\zeta(3)
      -{145\over 18} C_{\rm F}
      -{184\over 3}C_{\rm F}\zeta(3)
      +80C_{\rm F}\zeta(5)
\Biggr].
\label{EqExplicitLogScaleRatio}
\end{eqnarray}

Now the generalized Crewther relation takes the form
\begin{equation}
{1\over 3\sum_f Q_f^2} R_{e^+e^-}(s)
C_{\rm Bj}(Q^2)=1+\varepsilon_1(Q^2),
\label{EqBjCrewther}
\end{equation}
and
\begin{equation}
 {1\over 3\sum_f Q_f^2}  R_{e^+e^-}(s)
C_{\rm GLS}(Q^2)=1+\varepsilon_2(Q^2),
\label{EqGLSCrewther}
\end{equation}
with $\varepsilon_1$ and $\varepsilon_2$ defined above.

In what follows we will neglect the small values of $\varepsilon_1$
and $\varepsilon_2$  by simply  assuming that $C_{\rm Bj}\simeq
C_{\rm GLS},$ which is valid up to the NLO approximation of
perturbation theory and is almost valid at the NNLO (note that the
additional light-by-light-type contribution to the NNLO correction
to $C_{\rm GLS}$, which is included in the expression for
$\varepsilon_2(Q)$, is very small \cite{LarVerm}). The experimental
measurements of the $R$-ratio above the thresholds for the
production of $c\overline{c}$-bound states, together with the
theoretical fit performed in Ref. \cite{MattinglyStevenson}, provide
the constraint
\begin{equation}
{1\over 3\sum_f Q_f^2} R_{e^+e^-}(\sqrt
s=5.0~{\rm GeV})\simeq {3\over 10} (3.6\pm 0.1)=1.08\pm 0.03.
\end{equation}
and  thus
\begin{equation}
{\alpha_{R}^{\rm exp}(\sqrt s=5.0~{\rm GeV})
\over \pi}
 \simeq 0.08\pm 0.03.
\label{Eqalpha}
\end{equation}

As a consequence, from Eqs. (\ref{EqBjCrewther}) and
(\ref{EqGLSCrewther}) we  obtain the following estimate for $C_{\rm
Bjp}(Q)\simeq C_{\rm GLS}(Q)$:
\begin{equation}
C_{\rm Bj}(Q=12.33\pm 1.20~ {\rm GeV})
\simeq C_{\rm GLS}(Q=12.33\pm 1.20~ {\rm GeV})=0.926\mp 0.026,
\label{28}
\end{equation}
where the error bars for $Q$ and $C_{\rm Bj}(Q)$ are calculated using
the Eqs. (23) and  (24) respectively. The error bars correspond to
the uncertainties in the empirical value of $\alpha_R$ in Eq.
(27).\footnote{ Keeping the third significant digit after the
decimal point in these equations is an overestimate of the available
theoretical accuracy, but we will keep it nevertheless in order to
have better control of the real values of the second significant
digit in the r.h.s. of Eqs.(28)-(30).}

The corresponding expression for the effective coupling constants is
\begin{equation}
{\alpha_{g_1}^{\rm exp}(Q=12.33\pm 1.20~{\rm GeV})\over \pi}
\simeq
{\alpha_{\rm GLS}^{\rm exp}(Q=12.33\pm 1.20~{\rm GeV})\over \pi}
\simeq 0.074\pm 0.026;
\label{EqNumericalBjGLS}
\end{equation}

The predictions given above for $\alpha_{g_1}\simeq\alpha_{\rm GLS}$
are the predictions which can be tested experimentally.  The
recent measurements for the Gross-Llewellyn Smith sum rule are
performed only at relatively small values of $Q^2$
\cite{CCFRL1,CCFRL2}; however, one can use the results of the
theoretical extrapolation \cite{KS} of the experimental data presented
in \cite{CCFRQ} and turn to the domain of large values of $Q^2$.
Notice that the results of Ref.  \cite{KS} for the Gross-Llewellyn
Smith sum rule at $Q^2=3$ GeV$^2$ are in good agreement with the
well-determined experimental results (see
Refs. \cite{CCFRL1,CCFRL2}). Thus it is not difficult to extract
the value for $\alpha_{\rm GLS}(Q)/\pi$ from Table 1 of Ref.
\cite{KS}:
\begin{equation}
{\alpha_{\rm GLS}^{\rm extrapol}(Q=12.25~{\rm GeV})\over \pi}\simeq
0.093\pm0.042.
\end{equation}
This interval  overlaps with the result previously derived  in Eq.
(29): $0.074 \pm 0.026$. This gives empirical support for the
generalized Crewther relation derived in Eqs.(14)-(16) and in
Eqs.(24), (25). The relation of the results Eqs.(27)-(30) to the
commonly used language of the $\overline{\rm MS}$-scheme will be
presented elsewhere \cite{GKN}.

\bigskip
\leftline{{\bf 4. Discussions.}}
\bigskip

It is worthwhile to point out that all of the results presented here
are derived within the framework of perturbation theory and do not
involve the nonperturbative contributions to the Adler's function
$D(Q^2)$ \cite{SVZ} and the $R$-ratio, as well as to  the polarized
Bjorken and the Gross-Llewellyn Smith sum rules \cite{Jaffe,BBK}.
These nonperturbative contributions are expected to be significant
at small energies and momentum  transfer. In order to make these
contributions comparatively negligible, we have chosen relatively
large values for $s$ and $Q^2$ in our numerical study. Even at the
current level of understanding we can conclude that our numerical
analysis demonstrates reasonable agreement between the theoretical
predictions and experimental results. In order to put the analysis
of the experimental data for lower energies on more solid ground, it
will be necessary to understand whether there exist any
Crewther-type relations between non-perturbative order
$O(1/Q^4)$-corrections to the Adler's $D$-function \cite{SVZ} and
the order $O(1/Q^2)$ higher twist contributions to the
deep-inelastic sum rules \cite{Jaffe,BBK}.

The generalized Crewther relation written in the form of CSR
provides an important test of QCD. Since the Crewther formula
written in the form of the CSR relates one observable to another
observable, the predictions are independent of theoretical
conventions, such as the choice of renormalization scheme. It is
clearly  very interesting to test these fundamental self-consistency
relations between the polarized Bjorken sum rule or the
Gross-Llewellyn Smith sum rule and the $e^+e^-$-annihilation
$R$-ratio. Present data are consistent with the generalized Crewther
relations, but measurements at higher precision and energies will be
needed to decisively test these fundamental connections in QCD.

In order to check the consequence of the generalized Crewther
relation at a higher confidence level,  it will be necessary, first,
to reduce the experimental error of the measurement of $R_{e^+e^-}$
at $\sqrt{s}\approx 5~{\rm GeV}$ and, secondly, to have more precise
information on the value of the Gross-Llewellyn Smith sum rule at
$Q^2\approx 150~GeV^2$ or to measure the polarized Bjorken sum rule
at this momentum transfer. We hope that the first problem can be
attacked after the start of the operation of a $c-\tau$-factory.
Moreover, the possible future study of the deep inelastic scattering
with both polarized electron and proton beams can open the window
for the direct measurements of the polarized Bjorken sum rule at
high momentum transfer \cite{Blumlein}. Keeping this in mind we
propose direct measurements of the polarized Bjorken sum rule at the
scales $Q^2=150~{\rm GeV}^2$ and $Q^2=25~{\rm GeV}^2$. As was shown
in the previous section, the related measurements at the energy
scale $Q^2=150~{\rm GeV}^2$ can be helpful for checking the
generalized Crewther relation, written down in the form of the
single-scale CSR, derived in Eq. (14).

The direct measurements of the polarized Bjorken sum rule (or of the
Gross-Llewellyn Smith sum rule) at the scale $Q^2=25~{\rm GeV}^2$
can be useful for the study of the intriguing question whether the
experimental data can ``sense" the violation of the initial
conformal invariance caused by the renormalization procedure. In the
language of the Crewther relation this question can be reformulated
in the following manner: what will happen if we put the scales of
$R_{e^+e^-}$ and  the corresponding sum rules to be equal to each
other? Will the experimental data produce the unity on the r.h.s. of
Eq. (17), related to the conformal invariant limit, if we put
$s=|Q^2|$? Recall, that in this case, the theoretical expression for
the generalized Crewther relation will differ from the conformal
invariant result starting from the proportional to the well-known
factor $\beta(\alpha_s)/\alpha_s$ the $\alpha_s^2$-order corrections
\cite{BK}, which presumably reflects the violation of the conformal
symmetry by the procedure of renormalization \cite{Crewther,trace}
(for discussions see Ref. \cite{GK}). Notice, however, that the size
of the perturbative contribution proportional to the QCD $\beta$ -
function is rather small \cite{BK}.

\bigskip
\leftline{{\bf 5.  Acknowledgments}}
\bigskip

Two of us, S.J.B. and H.J.L., wish to thank M. Beneke, V. Braun,  P.
Huet, D.~M\"uller, and  J. Rathsman for helpful discussions. G.T.G.
and A.L.K. are grateful to V.A. Matveev, V.A. Rubakov and A.N.
Tavkhelidze for the useful conversations and the interest in the
problems related to this work. A.L.K. also wishes to thank
G.~Altarelli, D.J. Broadhurst and A.V. Sidorov for discussions. He
is also grateful to his colleagues from DESY (Hamburg), and
especially to P. M. Zerwas for  hospitality during the completing of
the final version of this work.

A preliminary version of this paper was reported at the XVIII
International Workshop on High Energy Physics and Field Theory
(26-30 June 1995, IHEP, Protvino, Russia) and at the Conference of
Nuclear Physics Division of the Russian Academy of Sciences
``Fundamental Interactions of Elementary Particles" (23-27 October
1995, ITEP, Moscow, Russia). The authors are grateful to the
organizers and the participants of these conferences for the
interest in this work and useful discussions.

The work of S.J.B. and H.J.L. is supported in part by the Department
of Energy, contracts DE--AC03--76SF00515 and DE--FG02--93ER--40792.
The work of A.L.K. is done within the project N 96-02-18897-a
submitted to the Russian Fund for Fundamental Research. The work of
G.T.G. was supported by the ISF grant  N6J000 and ICFPM graduate
student fellowship, grant 932492.

\end{document}